\documentclass[twocolumn,prd,noshowpacs,nofootinbib,amsmath,amssymb,superscriptaddress]{revtex4}
\usepackage{graphicx}
\usepackage{amssymb}
\usepackage{amsmath}
\usepackage{amsfonts}
\usepackage{latexsym}
\usepackage{hyperref}
\usepackage[all]{xy}
\usepackage{slashed}
\newcommand{\be}{\begin{eqnarray}}
\newcommand{\ee}{\end{eqnarray}}

\newcommand{\MeV}{~\mathrm{MeV}}
\newcommand{\GeV}{~\mathrm{GeV}}
\newcommand{\TeV}{~\mathrm{TeV}}
\newcommand{\cm}{~\mathrm{cm}}

\newcommand{\mZ}{M_{_Z}}
\newcommand{\mZp}{M_{_{Z'}}}
\newcommand{\mZone}{M_{_1}}
\newcommand{\mZtwo}{M_{_2}}

\newcommand{\cW}{c_{\rm _W}}
\newcommand{\sW}{s_{\rm _W}}
\newcommand{\cWp}{c_{\rm _{W'}}}
\newcommand{\sWp}{s_{\rm _{W'}}}

\newcommand{\JB}{J_{\rm _B}}
\newcommand{\JBp}{J_{\rm _{B'}}}
\newcommand{\JW}{J_{\rm _W}}
\newcommand{\JWp}{J_{\rm _{W'}}}

\newcommand{\JEM}{J_{\rm _{EM}}}
\newcommand{\JZ}{J_{\rm _Z}}
\newcommand{\JEMp}{J_{\rm _{EM'}}}
\newcommand{\JZp}{J_{\rm _{Z'}}}
\newcommand{\MQ}{M_{\rm mCP}}
\newcommand{\gsim}{\lower.7ex\hbox{$\;\stackrel{\textstyle>}{\sim}\;$}}
\newcommand{\lsim}{\lower.7ex\hbox{$\;\stackrel{\textstyle<}{\sim}\;$}}

\newcommand{\epsG}{\epsilon_{_\gamma}}
\newcommand{\epsZ}{\epsilon_{\rm _Z}}

\begin{document}

\title{A Milli-Window to Another World}
\author{Eder Izaguirre}
\affiliation{Perimeter Institute for Theoretical Physics, Waterloo, Ontario, Canada}
\author{Itay Yavin}
\affiliation{Perimeter Institute for Theoretical Physics, Waterloo, Ontario, Canada}
\affiliation{Department of Physics, McMaster University,  
Hamilton, ON, Canada}

\begin{abstract}
The kinetic mixing of the vector boson of hypercharge with the vector boson(s) associated with particle sectors beyond the Standard Model is one of the best motivated windows to new physics. The resulting phenomenology depends on whether the new vector boson is massive or massless. The phenomenology associated with the massive phase has received considerable attention in recent years with many theoretical explorations and new experimental efforts, while the massless phase is linked to the phenomenology of milli-charged particles. In this paper we introduce the more general case where the kinetic mixing is with a vector boson that is a linear combination of both a massive and a massless state (as hypercharge is in the Standard Model). We demonstrate that the general phase is only weakly constrained when the mass scale associated with it is above about 100 MeV. Finally, we show that a new dedicated experiment at the LHC, proposed recently in Ref.~\cite{Haas:2014dda}, can explore large parts of the parameter space in the mass range between 100 MeV and 100 GeV. In particular, it is uniquely sensitive to a new signature that only arises in the general phase.
\end{abstract}

\maketitle
\section{Introduction}

The existence of dark matter (DM) and the lessons from grand unified theories and string theory phenomenology suggest the presence of additional particle sector(s) beyond the Standard Model (SM). Such a sector, call it a dark sector (DS), may contain new matter and new short- and long-ranged forces, similar to those found in the SM. However, unless this DS interacts with our world in some way, this possibility would forever remain outside the purview of science. Quantum field theory guarantees that we interact with the DS through gravity~\cite{Weinberg:1964ew}, but that force is unfortunately too weak and too universal to illuminate the detailed structure of the DS. 

While it cannot provide us with assurances for additional interactions beyond gravity, quantum field theory does provide strong indications for the type of interactions we can hope for at accessible energy scales, namely marginal operators. One of the best motivated marginal operators connecting us with the DS is the kinetic mixing operator between our hypercharge field and an abelian vector field in the DS, 
\be
\label{eqn:def_lag}
\nonumber
\mathcal{L} &=& \mathcal{L}_{\rm SM} + \mathcal{L}_{\rm DS}  -\frac{\kappa}{2} B'_{\mu\nu}B^{\mu\nu}~.
\ee
Here $ \mathcal{L}_{\rm SM}$ and $ \mathcal{L}_{\rm DS}$ are the Lagrangians describing the particle content of the SM and the DS respectively, while $B_{\mu\nu}$ and $B'_{\mu\nu}$ are the field-strength of our hypercharge and the dark abelian vector field. The observable effects of the kinetic mixing operator depend on whether the dark vector field is massive or not. This in turn depends on the symmetry breaking pattern in the DS, which therefore leads to several distinct phases of the theory. 

In the first phase, which we will refer to as the {\it Okun phase} since it was first discussed by Okun in Ref.~\cite{Okun:1982xi}, the dark vector field $B'_\mu$ is massive. This phase is primarily characterized by the appearance of a massive vector boson with a coupling to the electromagnetic current suppressed by $\kappa$. The dark vector boson can thus be produced in electromagnetic processes and it can decay either into electrically charged SM matter or into matter in the DS. This phase of the theory has attracted renewed attention in recent years in connection with the dark matter problem~\cite{Pospelov:2007mp, ArkaniHamed:2008qn, Pospelov:2008jd}, which prompted numerous suggestions for new experiments and new searches devised to explore the parameter space associated with the Okun-phase. These include new searches at high-energy colliders~\cite{ArkaniHamed:2008qp,Baumgart:2009tn,Cheung:2009su}; new explorations at low-energy colliders looking for visible decays \cite{Pospelov:2008zw,Essig:2009nc, Bjorken:2009mm, Bjorken:1988as,Riordan:1987aw,Bross:1989mp,Davoudiasl:2012ig,Endo:2012hp,Babusci:2012cr,Adlarson:2013eza,Abrahamyan:2011gv,Merkel:2011ze,Reece:2009un,Aubert:2009cp} with several more expected to run in the next few years \cite{Essig:2010xa,Battaglieri:2014hga,Freytsis:2009bh,Wojtsekhowski:2009vz,Wojtsekhowski:2012zq,Beranek:2013yqa}; searches for invisible decays of the dark vector \cite{Adler:2004hp,Ablikim:2007ek,Artamonov:2009sz,deNiverville:2011it,Dharmapalan:2012xp,deNiverville:2012ij,Izaguirre:2013uxa,Essig:2013lka,Izaguirre:2014dua,Battaglieri:2014qoa,Izaguirre:2014bca,Batell:2014yra,Kahn:2014sra,Izaguirre:2015yja}; precision tests \cite{Pospelov:2008zw, Giudice:2012ms}; and effects in direct-detection experiments~\cite{Cheung:2009su,Finkbeiner:2009mi,Essig:2011nj,Essig:2012yx,Hochberg:2015pha}.

In the second phase, which we will refer to as the {\it Holdom phase}~\cite{Holdom:1985ag}, the dark vector field is massless. The principal physical effect associated with this phase is the appearance of milli-charged particles (mCPs). Any matter in the dark sector that couples to $B'_\mu$ now appears as a charged particle under electromagnetism with a charge proportional to $\kappa$, the kinetic mixing parameter\footnote{The word ``milli-" in this general context is somewhat imprecise since in general $\kappa$ may be much smaller than $10^{-3}$. Nevertheless, we will continue to refer to it as such since this value is physically motivated and lies within reach of the new experiments discussed in this paper.}. The existence of mCPs is constrained by a variety of observations including direct searches from accelerator experiments \cite{Prinz:1998ua, Davidson:2000hf, Badertscher:2006fm, CMS:2012xi} and indirect observations from astrophysical systems \cite{Davidson:1991si, Mohapatra:1990vq, Davidson:1993sj, Davidson:2000hf}, the cosmic microwave background \cite{Dubovsky:2003yn,Dolgov:2013una}, big-bang nucleosynthesis \cite{Vogel:2013raa}, and universe over-closure bounds \cite{Davidson:1991si}. In a recent paper~\cite{Haas:2014dda} we proposed a new experiment at the LHC with the potential of targeting the relatively unexplored part of the parameter space of mCPs with masses in the $100~\MeV - 100~\GeV$ range. Below we use the experimental proposal of Ref.~\cite{Haas:2014dda} as the basis for our projected sensitivity. 

In this paper we identify a third phase, which we refer to as the {\it mixed} phase,  where the dark boson $B'_\mu$ is in fact a linear combination of a massless state and a massive state. This is the analogue of hypercharge in the SM, which is a linear combination of the massless photon and the massive $Z$ boson. This third phase, being a mixture of the Okun and Holdom phases described above, is distinguished by the existence of a massive dark boson that can be produced directly through its coupling to the electromagnetic current, as well as by the appearance of mCPs. Thus, whereas in the Okun phase the decay of the dark boson into matter in the DS would simply constitute missing energy, in this third phase such decays would result in a pair of mCPs. This opens up a new production channel for mCPs beyond their direct production through the photon. A large part of this paper is devoted to a study of this new production channel in the context of the proposal of Ref.~\cite{Haas:2014dda} for a new experiment at the LHC to search for mCPs.

This third phase is subject to the same constraints as do the Okun and Holdom phases. In particular, the strong constraints on mCPs with masses below a few hundred MeVs are still applicable in this case, and so are the constraints on $\kappa$ associated with an extra massive dark boson below a few hundred MeV that couples to the electromagnetic current.  Indeed, it is precisely these constraints that force $\kappa$ to be extremely small ($\kappa \lesssim 10^{-8}$) in the case of ``mirror-world" scenarios where the lightest charged matter is the mirror-electron having the same mass as the SM electron\footnote{This remains true even in asymmetric mirror-world scenarios with scales larger by a factor of 20-30~\cite{Berezhiani:1995am,Mohapatra:2001sx}.}~\cite{Khlopov:1989fj,Foot:1995pa}. But the theoretical straightjacket associated with mirror-world scenarios seem ill-imposed: as we have learned from Twin Higgs theories~\cite{Chacko:2005pe,Chacko:2005un,Craig:2013fga} and more recently from the generalized framework of Orbifold Higgs~\cite{Craig:2014roa}, the mirroring of the gauge groups of the SM may be well-motivated without requiring a corresponding mirroring of the matter sector. Thus, in what follows we focus on the relatively unconstrained part of parameter space where both the massive dark boson and the mCPs have masses in the range of a few hundred MeVs to a few hundred GeVs, which permits a sizeable kinetic mixing parameter $\kappa \sim 10^{-3} - 10^{-2}$.

The remaining of this article proceeds as follows. First we present the model, an extension of the SM that includes a dark sector, and the kinetic mixing portal connecting the two in Sec.~\ref{sec:model}. Next, in Sec.~\ref{sec:existing constraints}, we review existing constraints on the parameter space of this framework. We discuss future prospects for probing the third phase of the model in Sec.~\ref{sec:future_probes}.

\section{Model}
\label{sec:model}

In this section we describe the model and derive the relevant mass eigenstates associated with the vector bosons and their interactions with the different SM and DS currents. 

\subsection{Mass eigenstates}

Besides the gauge field associated with hypercharge $B_\mu$, the SM also contains the gauge fields associated with the non-abelian group $SU(2)$, namely $W_{a\mu}$ with $a=1,2,3$. The gauge couplings are $g_1$ and $g_2$ corresponding to hypercharge and $SU(2)$, respectively. The spontaneous symmetry breaking associated with the Higgs boson results in the following mass eigenstates
\be
Z_1 &=& \cW W_3 - \sW B  \\
A_1 &=& \sW W_3 + \cW B 
\ee
where we suppressed the spacetime index on all fields for clarity, and defined (at tree-level) $\sW=\sin\theta_{\rm _W}=g_1/\sqrt{g_1^2+g_2^2}$, $\cW=\cos\theta_{\rm _W}=g_2/\sqrt{g_1^2+g_2^2}$, with  $\theta_{\rm _W}$ being the Weinberg angle. In the SM what we labeled as $Z_1$ is simply the $Z$-boson and $A_1$ is the photon. For now we keep the subscripts since, as we show below, the actual mass eigenstates in this theory are slightly different owing to the kinetic mixing in Eq.~(\ref{eqn:def_lag}). The vector field $Z_1^\mu$ has a mass term $\tfrac{1}{2}\mZone^2 Z_1^\mu Z_{1\mu}$, where the mass is related to the charged vector-boson mass through the usual relation $\mZone = M_{_{W^\pm}}/\cos\theta_{\rm _W}$.

Since we are interested in obtaining a similar structure in the DS, in what follows we will assume that in addition to the abelian gauge symmetry associated with the dark boson $B'_\mu$ there is at least one additional abelian or non-abelian group in the DS, with a gauge field $W'_\mu$ (whether it is part of a non-abelian group and carries another index as $W_{3\mu}$ does is irrelevant at this stage, and so we suppress this extra potential index). We denote the DS gauge couplings by $g_1'$ and $g_2'$, corresponding to the $B'_\mu$ and $W'_\mu$ gauge fields, respectively. We will further assume that spontaneous symmetry breaking in the DS leaves some linear combination of $B'_\mu$ and $W'_\mu$ massless while endowing the orthogonal linear combination with mass,
\be
Z_2 &=& \cWp W' - \sWp B'  \\
A_2 &=& \sWp W' + \cWp B' 
\ee 
This is the analogue to the SM combinations, and similarly we have $\sWp=\sin\theta_{\rm _{W'}}=g_1'/\sqrt{g_1^{\prime 2}+g_2^{\prime 2}}$, and $\cWp=\cos\theta_{\rm _{W'}}=g_2'/\sqrt{g_1^{\prime 2}+g_2^{\prime 2}}$, with  $\theta_{\rm _{W'}}$ being an analogue to the Weinberg angle in the SM. Here the vector field $Z_2^\mu$ has a mass term $\tfrac{1}{2}\mZone^2 Z_2^\mu Z_{2\mu}$.

In terms of these states, the kinetic mixing term can be written as,
\be
\nonumber
\kappa B_{\mu\nu}B^{\prime\mu\nu} = \kappa_{cc} A_1 A_2+ \kappa_{ss}Z_1 Z_2 - \kappa_{sc} Z_1 A_2 - \kappa_{cs}A_1 Z_2 \\
\ee
(here again we suppress the spacetime index on the right-hand side for clarity). We defined the coefficients $\kappa_{cc} = \kappa \cW \cWp$, $\kappa_{cs} = \kappa \cW \sWp$ and so on. One can remove the mixing exactly, but since we are interested in small $\kappa$ we do so only up to terms second order in $\kappa$ with a series of field redefinitions. We begin by removing the mixing between $A_1$ and $Z_2$ as well as $A_2$ and $Z_1$ through the simultaneous substitutions
\be
A_1 &\rightarrow& A_1 + \kappa_{cs} Z_2 \\
A_2 &\rightarrow& A_2 + \kappa_{sc} Z_1.
\ee
Here we mean that the right-hand side should be substituted for the left-hand side everywhere in the Lagrangian.
This results in mixing between $A_{1,2}$ and $Z_{1,2}$, which can be removed by
\be
A_1 &\rightarrow& A_1 - \kappa_{sc}\kappa_{cc} Z_1 \\
A_2 &\rightarrow& A_2 + \kappa_{cs}\kappa_{cc} Z_2
\ee
These transformations also induce a change in the coefficients of the kinetic terms for $Z_1$, and $Z_2$ which are now multiplied by $(1-\kappa_{sc}^2)$ and $(1-\kappa_{cs}^2)$, respectively (working to quadratic order in $\kappa$). These can be removed with the field redefinition,
\be
\label{eqn:firstZ1Z2rescaling}
Z_1 &\rightarrow& \frac{1}{\sqrt{1-\kappa_{sc}^2}} Z_1\\
Z_2 &\rightarrow& \frac{1}{\sqrt{1-\kappa_{cs}^2}} Z_2
\ee
 
Up to terms of order $\kappa^3$, which we neglect, the Lagrangian now only contains mixing between $A_1$ and $A_2$ as well as mixing between $Z_1$ and $Z_2$. The kinetic mixing between the massless states can be removed by a choice of basis as,
\be
\label{eqn:choice_of_massless_basis}
A_2 &\rightarrow& A' - \kappa_{cc} A\\
A_1 &\rightarrow& A. 
\ee
This final substitution chooses $A$ to be what we would call the photon, whereas $A'$ is the other massless state. The kinetic mixing between the massive states can be removed through the symmetric substitution, 
\be
Z_1 &\rightarrow& \tfrac{1}{\sqrt{2}} \left( Z_1 + Z_2 \right) \\
Z_2 &\rightarrow& \tfrac{1}{\sqrt{2}} \left( -Z_1 + Z_2 \right)
\ee
With these substitutions all the kinetic mixing terms are removed to the order we are working in, and a final re-scaling is necessary to end with canonical normalization for the fields,
\be
A &\rightarrow& \frac{1}{\sqrt{1-\kappa_{cc}^2}} A \\
Z_1 &\rightarrow& \frac{1}{\sqrt{1-\kappa_{ss}  }} Z_1 \\
Z_2 &\rightarrow& \frac{1}{\sqrt{1+\kappa_{ss}}} Z_2
\ee

The field redefinitions above result in mass mixing between $Z_1$ and $Z_2$. The mass matrix associated with the massive states is now given by,
\be
\frac{1}{2} \left(
\begin{array}{cc}
M_{11} &M_{12}\\
M_{21} & M_{22}
\end{array}
\right)
\ee
where the different entries are
\be
M_{11} &=& (1+\kappa_{ss})\left(\mZone^2+\mZtwo^2\right) \\ \nonumber &+& \kappa^2\left[\mZone^2 \sW^2 + \mZtwo^2 \sWp^2 \right] \\
M_{12} &=& M_{21} = \left(\mZone^2 - \mZtwo^2\right) \\ \nonumber &+&  \kappa^2 \left[ \mZone^2\left(1-\tfrac{1}{2} \sWp^2 \right) \sW^2 - \mZtwo^2\left(1-\tfrac{1}{2} \sW^2 \right) \sWp^2 \right] \\
M_{22} &=&  (1-\kappa_{ss})\left(\mZone^2+\mZtwo^2\right) \\ \nonumber &+& \kappa^2\left[\mZone^2 \sW^2 + \mZtwo^2 \sWp^2 \right]
\ee
up to terms of order $\kappa^3$. This mass matrix can be diagonalized in perturbation theory in $\kappa$ to yield the final mass eigenvalues, 
\be
\label{eq:mz}
\mZ^2 = \mZone^2\left( 1+ \kappa^2\sW^2\frac{\mZone^2 - \mZtwo^2\cWp^2}{\mZone^2 - \mZtwo^2} \right) \\
\label{eq:mzp}
\mZp^2 = \mZtwo^2\left(1+ \kappa^2\sWp^2\frac{\mZtwo^2 - \mZone^2\cW^2}{\mZtwo^2 - \mZone^2} \right)
\ee
with $\mZone = M_{_{W^\pm}}/\cW$. The order $\kappa^2$ shift in the $Z$-boson mass represents a general and important constraint on this model as we shall discuss below~\cite{Hook:2010tw}. The corresponding mass eigenstates given in terms of the $Z_1$ and $Z_2$ states are,
\be
\hspace{-3mm}
\left(
\begin{array}{c}
Z \\
Z' 
\end{array}
\right) 
=\left(
\begin{array}{cc}
\cos \phi & -\sin\phi \\
\sin\phi & \cos\phi
\end{array}
\right) 
\left(
\begin{array}{cc}
\tfrac{1}{\sqrt{2}} &\tfrac{1}{\sqrt{2}} \\
-\tfrac{1}{\sqrt{2}} & \tfrac{1}{\sqrt{2}}
\end{array}
\right) 
\left(
\begin{array}{c}
Z_1 \\
Z_2 
\end{array}
\right) 
\ee
with the mixing angle given by
\be
\sin\phi &=& \tfrac{1}{2}\kappa_{ss} \frac{\mZone^2+\mZtwo^2}{\mZone^2-\mZtwo^2} \\
\cos\phi &=& 1- \frac{1}{2} \bigg(\tfrac{1}{2}\kappa_{ss} \frac{\mZone^2+\mZtwo^2}{\mZone^2-\mZtwo^2} \bigg)^2
\ee
to leading order in $\kappa$. This completes the diagonalization and results in two massive states $Z$ and $Z'$, as well as two massless states $A$ and $A'$. We now move on to the interactions of these vector bosons with the different SM and DS currents. 

\subsection{Interactions}

We write the interactions in the original gauge basis as follows,
\be
\hspace{-8mm} \mathcal{L} \supset g_1 B_\mu \JB^\mu + g_2 W_{3\mu} \JW^\mu + g'_1 B'_\mu \JBp^\mu + g'_2 W'_{3\mu} \JWp^\mu
\ee
Here $ \JB^\mu$ and $\JW^\mu$ are the familiar $U_{_Y}(1)$ and $SU_{_W}(2)$ currents of the SM containing leptons and quarks, whereas $ \JBp^\mu$ and $ \JWp^\mu$ are the corresponding currents in the DS, to whose content we return to later. 
The usual electromagnetic current and $Z$ current are the linear combinations,
\be
\JEM &=& \JB + \JW \\
\JZ &=& -\sW^2 \JB + \cW^2 \JW
\ee
and similarly for the DS currents. 

After removing the kinetic mixing through the field redefinitions above and rotating to the mass basis,  the different mass eigenstates couple as follows. Our photon couples as,
\be
\label{eqn:couplings_of_photon}
\mathcal{L} \supset A \cdot \Bigg[ e\left(1+\tfrac{1}{2} \kappa_{cc}^2 \right)  \JEM - \kappa_{cc} e'  \JEMp \Bigg]~,
\ee
where $e=g_1\cW$ is the usual definition of the electromagnetic coupling. Here we suppressed the spacetime indices on the gauge field and the currents, as we do below as well. 
These interactions of the photon have a simple interpretation: other than a rescaling of what we call electric charge by $1+\tfrac{1}{2} \kappa_{cc}^2$, the photon now also couples to the DS's charged particles with an effective electromagnetic charge of $\kappa_{cc}e'$ - hence, milli-charge. This corresponds to the appearance of mCPs expected in the Holdom phase, which is associated with the existence of the second massless state, $A'$. 

Next, what we know as the $Z$-boson now couples as,
\be
\label{eqn:Zcouplings}
\nonumber
\mathcal{L} &\supset&  Z \cdot  \Bigg[  \frac{g_2}{\cos\theta_{\rm _W}}(1+\xi_{_{ZZ}} ) \JZ - e \xi_{_{ZE}} \JEM \\ \nonumber
~\\ 
&~& \hspace{10mm} +e' \xi_{_{ZE'}} \JEMp  - \frac{g_2'}{\cos\theta_{\rm _{W'}}} \xi_{_{ZZ'}} \JZp \Bigg] 
\ee
with
\be
\xi_{_{ZZ}} &=& \tfrac{1}{2} \kappa^2 \sW^2\left( 1 - \frac{\mZtwo^4\sWp^2}{\left(\mZone^2-\mZtwo^2\right)^2}\right) \\
\xi_{_{ZE}} &=& \kappa^2 \cW \sW \left( \frac{\mZone^2 - \mZtwo^2\cWp^2}{\mZone^2 - \mZtwo^2} \right) \\
\xi_{_{ZE'}} &=& \kappa_{sc} \\
\xi_{_{ZZ'}} &=& \kappa_{ss}\left(\frac{\mZone^2}{\mZone^2-\mZtwo^2}\right).
\ee
Here, since we are only working to order $\kappa^2$ one can replace $\mZone$ and $\mZtwo$ by the physical masses $\mZ$ and $\mZp$Z. The interactions in Eq.~(\ref{eqn:Zcouplings}) again have a fairly straightforward interpretation: the first line shows that the standard $Z$ coupling is rescaled by $1+ \xi_{_{ZZ}}$ and the current receives a contribution of order $\kappa^2$ from the electromagnetic current. This would effect forward-backward observables and other observables at the $Z$ pole. The second line of Eq.~(\ref{eqn:Zcouplings})  shows that the $Z$ now also couples to the DS currents at order $\kappa$. Thus, $Z$ rare decays would contribute to the production of mCPs. 

Next are the DS vector bosons. Due to the choice of basis we made in Eq.~(\ref{eqn:choice_of_massless_basis}) the massless state $A'$ entirely decouples from the standard model with interactions only with the mCPs,
\be
\mathcal{L} &\supset& e' A' \cdot \JEMp~,
\ee
where $e' = g_1'\cWp$. Therefore, the massless dark photon is essentially decoupled from the SM, but it does potentially play a role in the relic abundance of the mCPs (see below). It may also be of importance in enhanced radiation effects in the DS if $e'$ is sizable. 

Finally, the interactions of the massive state $Z'$ are,
\be
\label{eqn:Zpcouplings}
\nonumber
\mathcal{L} \supset&~& Z' \cdot \Bigg[  \frac{g'_2}{\cos\theta_{\rm _{W'}}} \left(1+ \xi_{_{Z'Z'}} \right)  \JZp - e' \xi_{_{Z'E'}} \JEMp \\ \nonumber
~\\ 
&~& \hspace{10mm} +e~ \xi_{_{Z'E}} \JEM + \frac{g_2}{\cos\theta_{\rm _W}} \xi_{_{Z'Z}} ~ \JZ \Bigg] 
\ee
with
\be
\xi_{_{Z'Z'}} &=& \tfrac{1}{2} \kappa^2 \sWp^2\left( 1 - \frac{\mZone^4\sW^2}{\left(\mZone^2-\mZtwo^2\right)^2}\right) \\
\xi_{_{Z'E'}} &=& \kappa^2 \cWp \sWp \left( \frac{\mZone^2\cW^2 - \mZtwo^2}{\mZone^2 - \mZtwo^2}\right) \\
\xi_{_{Z'E}} &=& \kappa_{cs}  \\
\xi_{_{Z'Z}} &=& \kappa_{ss} \left( \frac{\mZtwo^2}{\mZone^2-\mZtwo^2}\right)
\ee
The second line of Eq.~(\ref{eqn:Zpcouplings}) is the mixing of the massive $Z'$ state with the SM, familiar from the Okun phase of the theory. If there are light charged particles in the DS to which the $Z'$ can decay to, then the first line of Eq.~(\ref{eqn:Zpcouplings}) is more important in our case as it would result in the production of mCPs through $Z'$ decay. Since we are not sensitive to subdominant effects, we can neglect the $\kappa^2$ terms in the coupling of $Z'$ to mCPs, which is then simply 
\be
 \mathcal{L} \supset \frac{g_2'}{\cos\theta_{\rm _W'}} \JZp \cdot Z'
\ee

\section{Existing Constraints}
\label{sec:existing constraints}
We now turn our discussion to the existing constraints on the model presented in Sec.~\ref{sec:model}. The free parameters of the model are the mass and dark-charge of the lightest\footnote{We focus on the lightest mCP because, unless there exist heavier mCPs with far larger dark-charges, the lightest mCP would be the one particle primarily produced and hence dominate the constraints.} mCP, $\MQ$ and $e'$, respectively;  the massive dark boson mass $\mZp$; the dark Weinberg angle $\theta_{W'}$; and the kinetic mixing parameter $\kappa$. For clarity, below we discuss separately the known constraints on the pure Holdom phase ($\theta_{W'}=0$) and on the pure Okun phase ($\theta_{W'}=\pi/2$). The experimental bounds associated with these two limiting cases become more or less relevant in the general mixed phase depending on how close $\theta_{W'}$ is to $0$ or $\pi/2$. 

In the Holdom phase the constraints come from previous searches for mCPs and therefore probe the parameter space spanned by the mass of the mCP, $\MQ$, and its electric charge given by,
\be
\label{eqn:milli_charge}
\epsG e\equiv \kappa  e' \cW \cWp~.
\ee
Here we have defined the fractional charge $\epsG$ in units of the electron's charge $e$. 

On the other hand, in the Okun phase, the constrained parameter space is spanned by the mass of the dark-boson $\mZp$ and its coupling to the SM. This coupling, at least in the limit $\mZp \ll \mZ$, is simply given by,
\be
\epsZ e \equiv \kappa e ~\cW \sWp
 \ee 
 where we have defined the fractional Weak charge $\epsilon_Z$ in units of $e$. For the mixed phase, laboratory probes will depend on the product of $\epsG \epsilon_Z$, as we will discuss below. One exception comes from electroweak precision tests which are primarily sensitive to the combination $\kappa\sin\theta_W$ and depend only mildly on $\mZp$ in the limit of  $\mZp \ll \mZ$ (see Eq.~\ref{eq:mz}).

\subsection{Constraints on the Holdom Phase}
\label{sec:massless constraints}

 Existing constraints on mCPs can be categorized as either coming from indirect probes (cosmology and astrophysics) or from direct searches at laboratory experiments. The former tend to be considerably more model-dependent than the latter, and can be weakened through additional matter in the DS which mCPs can annihilate into. In the range of couplings that this article focuses on, the mCP particles can be copiously produced in the early Universe. However, as the Universe evolves, the density of mCPs is depleted through pair annihilation in the DS. This process can give a large enough rate to avoid the different bounds on the relic abundance of the mCP~\cite{Davidson:1991si, Dubovsky:2003yn, Dolgov:2013una,Vogel:2013raa}. Another indirect constraint comes from the number of relativistic degrees of freedom, known as $N_{\rm eff}$, however since the dark massless photon decouples together with the mCP it will be much colder than the rest of the SM after entropy injection at later times. Moreover, the contribution to $N_{\rm eff}$ scales like the fourth power of the temperature, therefore this contribution is negligible when decoupling happens above a GeV or so~\cite{Jaeckel:2010ni, Brust:2013ova,Vogel:2013raa}.

Laboratory experiments place direct limits on mCPs for $10^{-5} \lsim \epsG \lsim 10^{-1}$ below about $\MQ < 300\GeV$. These are the result of a dedicated experiment that searched for mCPs at SLAC \cite{Prinz:1998ua}; accelerator experiments consisting of searches at beam-dump experiments,  free-quark searches, trident process searches, bounds from the invisible width of the $Z$ as well as direct searches for fractionally-charged particles at LEP \cite{Davidson:2000hf}; and decays of ortho-positronium \cite{Badertscher:2006fm}. Recently, CMS performed a search that excluded fractionally charged particles with charge $\pm e/3$ for $\MQ < 140\GeV$ and particles with charge $\pm2e/3$ for $\MQ < 310\GeV$~\cite{CMS:2012xi}. The least explored part of the parameter space is mCPs in the range $100\MeV < \MQ < 200 \GeV$, where $\epsG \lsim 10^{-1}$ is only barely explored. This target of opportunity motivated the proposal in Ref.~\cite{Haas:2014dda}, which we summarize in Sec.~\ref{sec:future_probes}.

Finally, we note that there exist strong constraints on the flux of mCPs and on their presence in bulk matter (see ref.~\cite{Perl:2009zz} for a detailed review). For example, in a recent publication~\cite{Agnese:2014vxh}, the CDMS collaboration reported on strong new limits on the flux of mCPs with charge between $e/200$ and $e/6$. However, such constraints cannot be directly applied to the model parameters since they require knowledge of the flux of mCPs on earth, which may be very small --- production in cosmic rays is negligible and relics from the early universe may be absent altogether either because of washout or because they were expelled by the magnetic fields of the galaxy. Thus it is challenging to interpret the null results coming from these experiments in the context of the theory we discuss. 

\subsection{Constraints on the Okun Phase}
\label{sec:massive constraints}

The Okun phase has been explored by numerous previous experiments, motivated by the case when some matter component in the DS is the dark matter\footnote{In the model presented here, however, only matter in the DS that is charged under the $Z'$, but not under the $A'$ (the analogue of neutrinos) can potentially be the DM, see more in section~\ref{sec:conclusions}. This is so because DS matter that is charged under the $A'$ becomes millicharged and there exist strong constraints on electrically-charged relics~\cite{McDermott:2010pa,Dvorkin:2013cea}.} of the Universe  \cite{Pospelov:2008zw,Essig:2009nc, Bjorken:2009mm, Bjorken:1988as,Riordan:1987aw,Bross:1989mp,Davoudiasl:2012ig,Endo:2012hp,Babusci:2012cr,Adlarson:2013eza,Abrahamyan:2011gv,Merkel:2011ze,Reece:2009un,Aubert:2009cp,Adler:2004hp,Ablikim:2007ek,Artamonov:2009sz,deNiverville:2011it,Dharmapalan:2012xp,deNiverville:2012ij,Izaguirre:2013uxa,Essig:2013lka,D'Agnolo:2015koa}. One must make a distinction between searches for \emph{visible} decays of the $Z'$ into SM particles, and searches for  \emph{invisible} decays into the DS. The former rely on the assumption that the $Z'$ decays 100\% into the SM, but in the presence of matter in the DS with $\mathcal{O}(1)$ couplings to the $Z'$, the decays of the latter into the SM are sub-dominant. The various experimental results we discuss below will primarily constrain the production strength of the $Z'$, which we parametrize by $\epsZ e$.

Visible searches have constrained the Okun phase primarily by looking for production of the $Z'$ and its subsequent decay into dilepton pairs $\ell^+\ell^-$. These become relevant in the case when decays into the DS are kinematically forbidden. For $\epsZ$ near $10^{-3}$, previous beam dump experiments place the strongest limits for $2 m_e \lsim \mZp \lsim 100 \MeV$, with planned fixed-target experiments expected to extend coverage to smaller couplings for $\mZp\lsim 500 \MeV$ (see Ref.~\cite{Essig:2013lka} for an exhaustive review of prospects for future probes). For masses above $\mZp \sim 1\GeV$, B-factories become important and have placed the strongest limits to date through searches for the $e^{+}e^{-}\rightarrow X+(Z' \rightarrow \ell^{+}\ell^{-})$ final state. Moreover, searches at the LHC for dilepton resonances prove an effective probe for $\mZp$ below the $Z$-mass \cite{Hoenig:2014dsa}.

Precision electroweak tests further restrict the parameter space for higher mass $Z'$ \cite{Hook:2010tw,Curtin:2014cca} and are independent of the way in which the $Z'$ decays. In particular, the shift in the mass of the $Z$ boson due to mixing with the $Z'$, Eq.~(\ref{eq:mz}) above,  places a limit on $\kappa$ of $\kappa < 10^{-2}$ for masses below the $Z$ mass. At the $Z$-pole, the bound on $\kappa$ improves to $\sim 10^{-3}$ due to the constraints on the modifications of the $Z$-coupling to SM fermions induced by the $Z'$.

In the regime where $\mZp$ is sufficiently large to allow for a direct decay back to DS particles, these invisible decays of the $Z'$ into the DS dominate for the $\kappa$ of interest to this paper, and the above limits from visible searches become irrelevant as the decays of the $Z'$ into the SM are further suppressed by $\kappa^2$. Proton and electron beam-dump experiments have established the strongest limits to date for invisibly-decaying $Z'$ below a few hundred MeV and mixings near $10^{-3}$. Moreover, for masses outside the reach of fixed-target experiments, the $Z'$ is best constrained by B-factories experiments through a $\gamma+$invisible final state search \cite{Izaguirre:2013uxa, Essig:2013vha}. 

By far the least explored part of the parameter space of our model is in the $\mZp > 100 \MeV$, $\epsZ \sim 10^{-3}$ regime. Recent proposals to target the sub-GeV---few-GeV part of this parameter space could sharply test the minimal models of light DM in the Okun phase \cite{Essig:2013vha,Chavarria:2014ika,Cushman:2013zza,Gerbier:2014jwa,Izaguirre:2014bca,Izaguirre:2015yja}, however a vast window into the other phases of the DS will remain untested by existing or planned experimental efforts. We now turn our discussion to a potentially powerful new probe of a new signature that only arises in the mixed phase of the model in Sec.~\ref{sec:model}.

\section{Probing the mixed phase at the LHC}
\label{sec:future_probes}

\begin{figure}[t]
\centering
\includegraphics[width=0.48 \textwidth ]{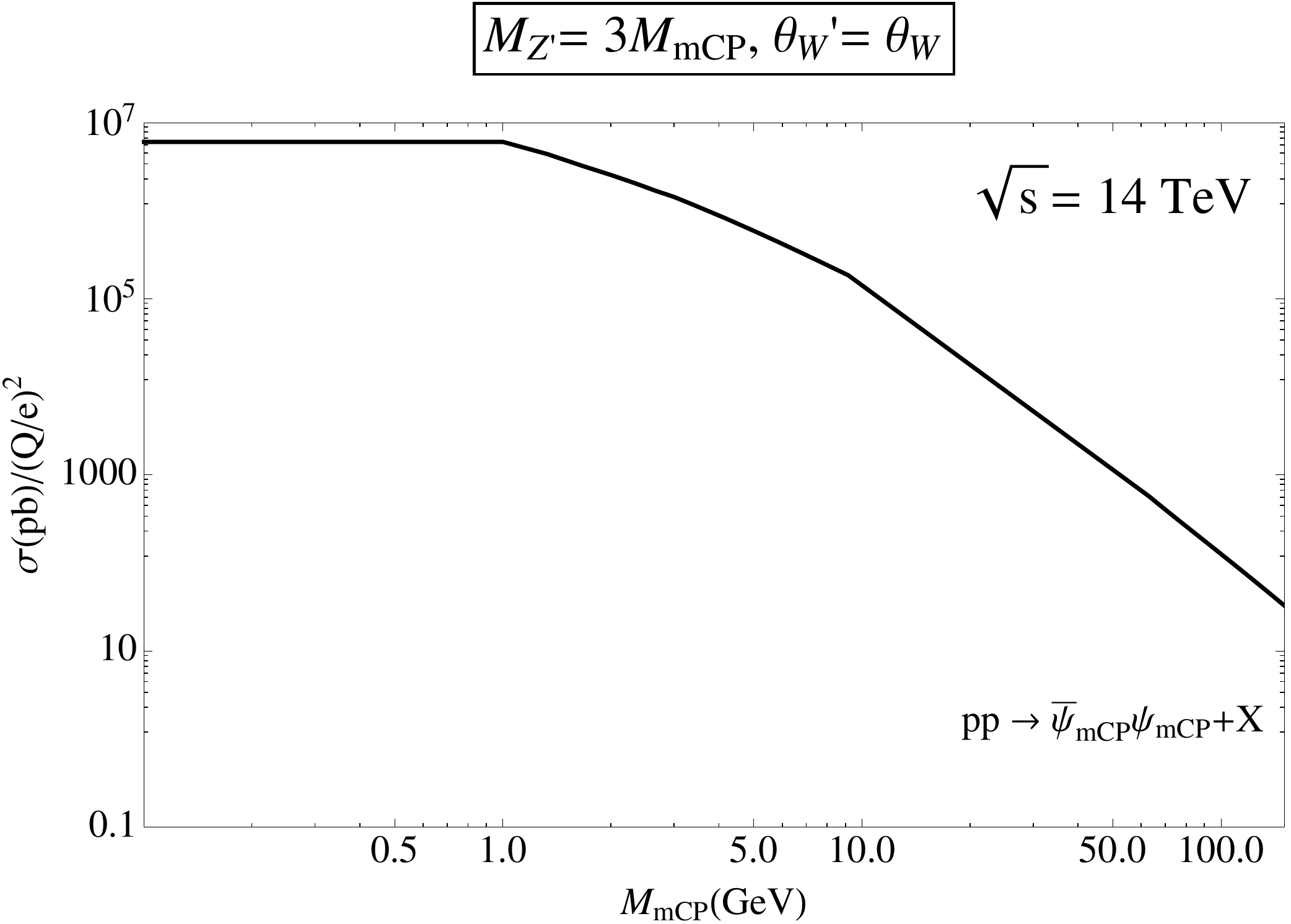}\\ 
\caption{Production cross section of mCPs at the LHC through the reaction $pp \rightarrow \bar\psi_{\rm mCP} \psi_{\rm mCP}+X$ with $M_{Z'}=3 M_{\rm mCP}$ and $\theta_{W'}=\theta_W$. The production of mCPs receives contributions from Drell-Yan like production, as well as from the decay of quarkonia. We impose a cut $M_{\bar\psi_{\rm mCP} \psi_{\rm{mCP}}}> 2\GeV$ to make sure we are in a physical region of the parton distribution functions. We conservatively fix the cross-section below  $M_{\rm mCP} < 1\GeV$. }
\label{fig:xsection}
\end{figure}

In this section we discuss a new signature of the DS --- absent in the Holdom and Okun phases --- that only emerges in the mixed phase, and we show that a recent proposal for a new experiment at the LHC is uniquely suited to look for this signature. The LHC is an obvious choice for probing the $m_{Z'} > 1 \GeV$ regime through $Z'$ decays into the mCP, as the $Z'$ could be copiously produced via processes analogous to the production of SM EW bosons. In particular, in the $pp\rightarrow (Z'\rightarrow \bar\psi_{\rm mCP} \psi_{\rm mCP}) + X$ reaction, the mCPs would register as missing energy at either ATLAS and CMS. However, the sensitivity of jet/photon and missing energy searches to this reaction is severely limited by large irreducible backgrounds from $Z\rightarrow \bar\nu\nu +$jets and from instrumental backgrounds from mis-measured jets \cite{Haas:2014dda}. The proposal from Ref.~\cite{Haas:2014dda} focused on the production and detection of mCPs, using the Holdom phase as a simple framework for generating non-quantized electromagnetic charges for beyond the SM matter. In fact, the LHC could also be a factory of mCPs originating from resonant-production of the massive $Z'$. The production of mCPs through the decay of the $Z'$ is a process unique to the mixed phase of the DS, and to our knowledge has not been considered before in the literature. Fig.~\ref{fig:xsection} shows the sizeable inclusive production cross section for the process $pp\rightarrow \bar\psi_{\rm mCP} \psi_{\rm mCP}$ at the $\sqrt{s}=14\TeV$ LHC, which we calculate with \texttt{Madgraph 5} \cite{Alwall:2014hca} and \texttt{MADONIA} \cite{Artoisenet:2007qm} -- the latter for mCPs from resonant production and decay of quarkonia.

\begin{figure}[t]
\centering
\includegraphics[width=0.48 \textwidth ]{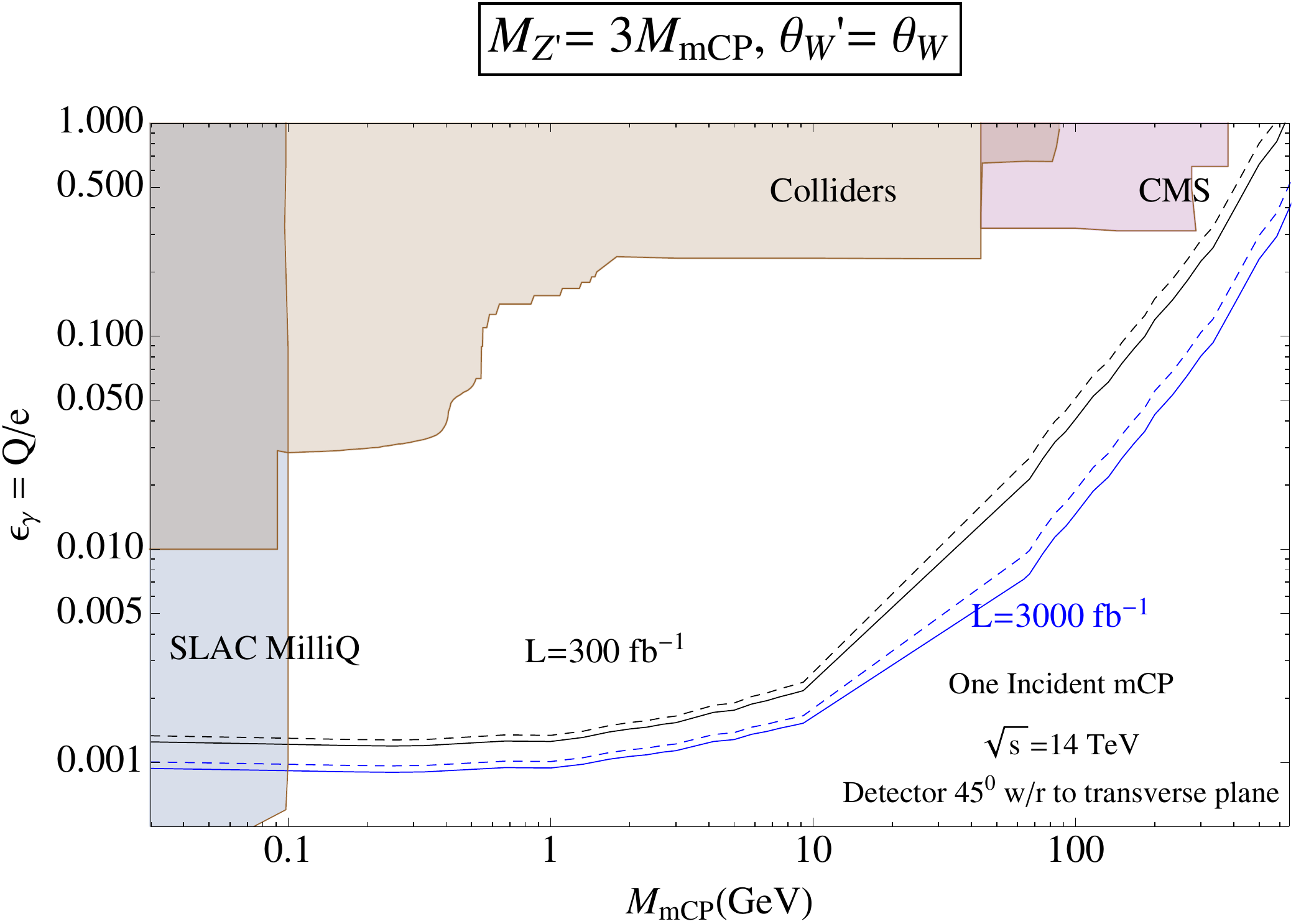}\\ 
\includegraphics[width=0.48 \textwidth ]{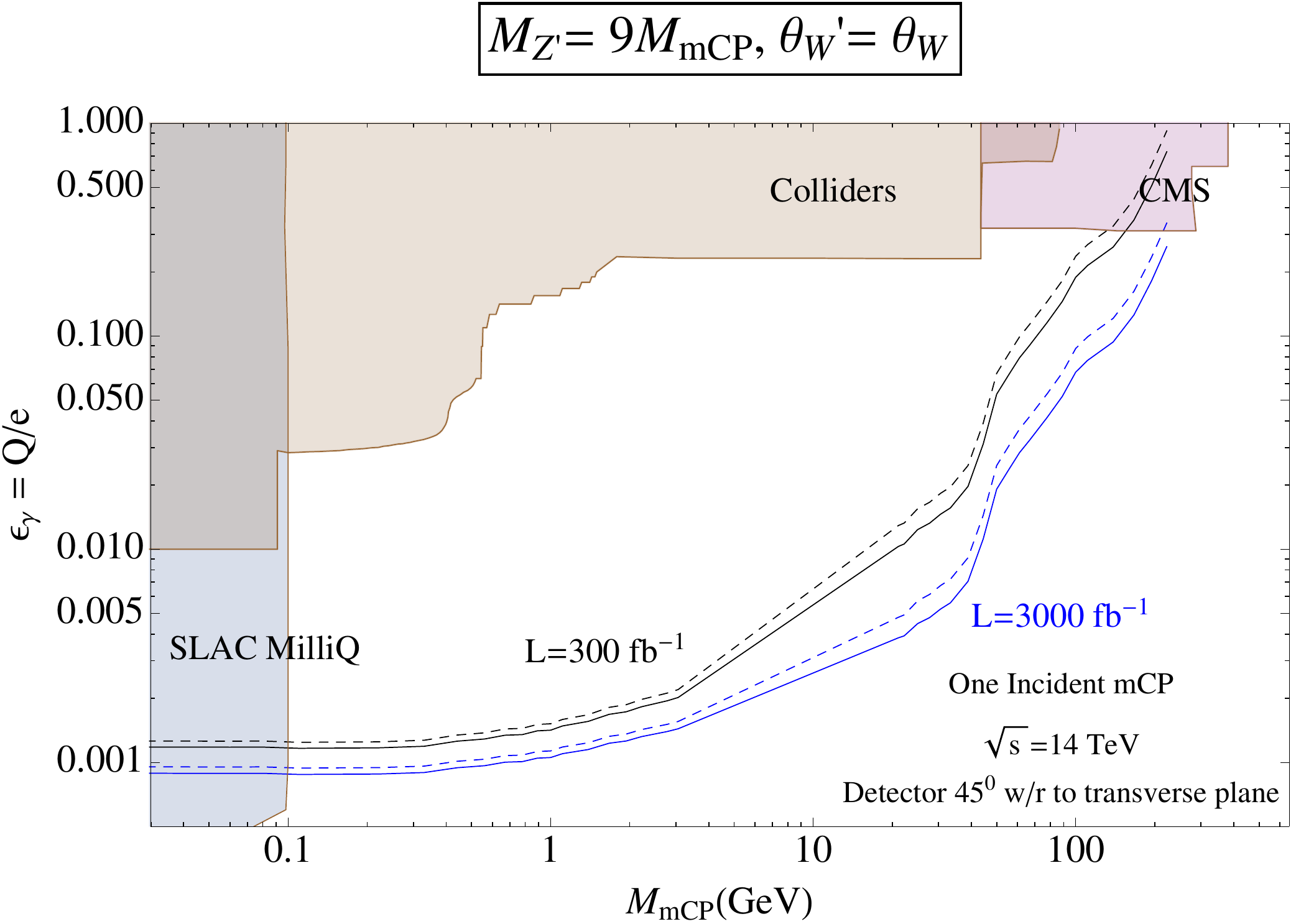}\\
\caption{Existing direct constraints on the mCP of the DS. In the regime $100\MeV<\MQ<200\GeV$ existing constraints come from direct searches at colliders as well as the bound from the invisible width of the $Z$ from LEP. In black (blue) we show the estimated sensitivity of the experiment at the LHC proposed in Ref.~\cite{Haas:2014dda} for 300 fb$^{-1}$  (3000 fb$^{-1}$). The dashed (solid) lines denote the $3\sigma$ sensitivity ($2\sigma$ exclusion). The top (bottom) panel shows the estimated sensitivity for the case $M_{Z'}=3 M_{\rm mCP}$ ($M_{Z'}=9 M_{\rm mCP}$) and $\theta_{W'}=\theta_W$. Note that we omit indirect constraints from cosmology since these are of a more model-dependent nature. In particular, the extraction of $N_{\rm eff}$ at the CMB disfavors masses between $0.1\GeV < \MQ < 1\GeV$, and the constraint from the anisotropies of the CMB disfavors the $\MQ > 230\GeV$ regime.}
\label{fig:moneyplotlowthetap}
\end{figure}

In what follows we summarize the details of the proposal from Ref.~\cite{Haas:2014dda} to look for minimally ionizing mCPs produced at either of the LHC interaction points by placing a small-scale detector in their vicinity. Following production at the interaction point (IP), mCPs would escape and travel unimpeded through rock. Ref.~\cite{Haas:2014dda} proposed placing a $\sim 1 \rm{m}^3$ detector some $\sim 20$ m  away from the IP and outside of ATLAS or CMS pointing the small detector towards the IP. An example of the kind of detector needed to efficiently search for mCPs with electric charges much smaller than $e$ was that used by the dedicated milli-charge experiment at SLAC \cite{Prinz:1998ua}. It consisted of a (plastic) scintillator in a quiet environment, whose purpose was to detect minimally ionizing signals. Indeed, the mCP signal to detect would be $\mathcal{O}(1)$ photoelectrons (PE) from a single excitation or ionization as the mCPs pass through the scintillating material and interact electromagnetically. Such a signal would be dominated by dark noise, typically at the level of 1 KHz \cite{pmthandbook}, and indeed this was one of the limiting factors in the sensitivity of the SLAC experiment. However, this difficulty can be overcome by using not one, but a series of back-to-back scintillator layers each looking for single-few PE signals coincident in time. The telescopic orientation of the detector combined with the directionality associated with a coincident signal in back-to-back layers could prove extremely useful in mitigating environmental backgrounds. Ref.~\cite{Haas:2014dda} estimated that a telescopic-coincident configuration of three layers, each a $1\rm{m}\times1\rm{m}\times1.4\rm{m}$ block, could reduce the dark noise backgrounds to a negligible level. Moreover, based on Ref.~\cite{Prinz:1998ua}, the results in Fig.~\ref{fig:moneyplotlowthetap} and Fig.~\ref{fig:twovsone} assume a 10\% probability to detect a signal PE, and that the probability of detection is Poisson-distributed with a mean 
\be
N_{\rm PE}=\left(\frac{\epsG}{2\times10^{-3}}\right)^2.
\label{eq:averagepe}
\ee

The experimental proposal discussed above could be sensitive to a large subset of the parameter space of the mixed phase that still remains viable. For concreteness, we consider the case $\theta_{W'}=\theta_W$, although this argument extends to the small $\theta_{W'}$ regime in general ({\it i.e.,} a mostly Holdom-like mixed phase). Fig.~\ref{fig:moneyplotlowthetap} shows the existing constraints on the $\MQ$ and $\epsG$ plane, for two different values of the ratio $\mZp/\MQ$. Also shown in Fig.~\ref{fig:moneyplotlowthetap} are the projections from the proposed experiment for mCPs at the LHC~\cite{Haas:2014dda}. The production of mCPs we focus on in this paper is primarily through the $Z'$ and is therefore sensitive to both $\epsZ$ (production of the $Z'$) and $\epsG$ (detection of the mCP). The production of the $Z'$ is limited by the constraints on $\epsZ$ from searches for its invisible decay, which are still applicable in the mixed phase. In particular, for $\mZp< 8 \GeV$, the model we consider is subject to the strong constraints from B-factories, which set an upper bound of $\epsZ/\cos\theta_W < 10^{-3}$. This limit in turn implies $\kappa < 10^{-3}$ for $\theta_{W'}=\theta_W$. The translation of this limit to a limit on $\epsG$ depends on the unknown dark charge $e'$. It in turn can be as large as $e' \approx 9e$ without introducing a Landau pole in the running of the dark $\beta$ function at dangerously low energies. If we require perturbation theory to be applicable we find the limit $\epsG\approx 4\times10^{-3}$ for $\theta_{W'}=\theta_W$ for $\mZp< 8 \GeV$. Even such small $\epsG$ could still be within the reach of the LHC mCP experiment proposal as shown by Fig.~\ref{fig:moneyplotlowthetap}. Above the kinematic reach of B-factories, in the entire range $\mZp > 8 \GeV$ and not too close to the Z-pole, the main constraint is the precise measurement of $\mZ$, which only limits $\kappa$ to be less than $10^{-2}$. In this case, $\epsG$ can be as large as $\rm{few}\times10^{-2}$, which lies in a viable part of the parameter space of the DS mixed-phase and well within reach of the LHC mCP experimental proposal.

Finally, we note that certain regimes of the mixed phase could remain challenging to probe by existing and planned experiments, as well as the proposal from Ref.~\cite{Haas:2014dda}. In particular, the limit of large $\theta_{W'}$ ({\it i.e.,} a mostly Okun-like theory) is challenging. For example, consider the case $\sin^2\theta_{W'} = 0.75$. In that scenario, for $\mZp < 8\GeV$ B-factories results imply $\kappa < 10^{-3}$. Setting $\kappa$ to that value in turn means that the largest $\epsG$ that could be generated without encountering a Landau Pole at low energies would be $\epsG \sim 1\times10^{-3}$, which may be too small even for the proposal from Ref.~\cite{Haas:2014dda} to be sensitive to.

\subsection{Unique Signatures of the Mixed Phase}
\label{subsec:otherhandles}

\begin{figure}[t!]
\centering
\includegraphics[width=0.48 \textwidth ]{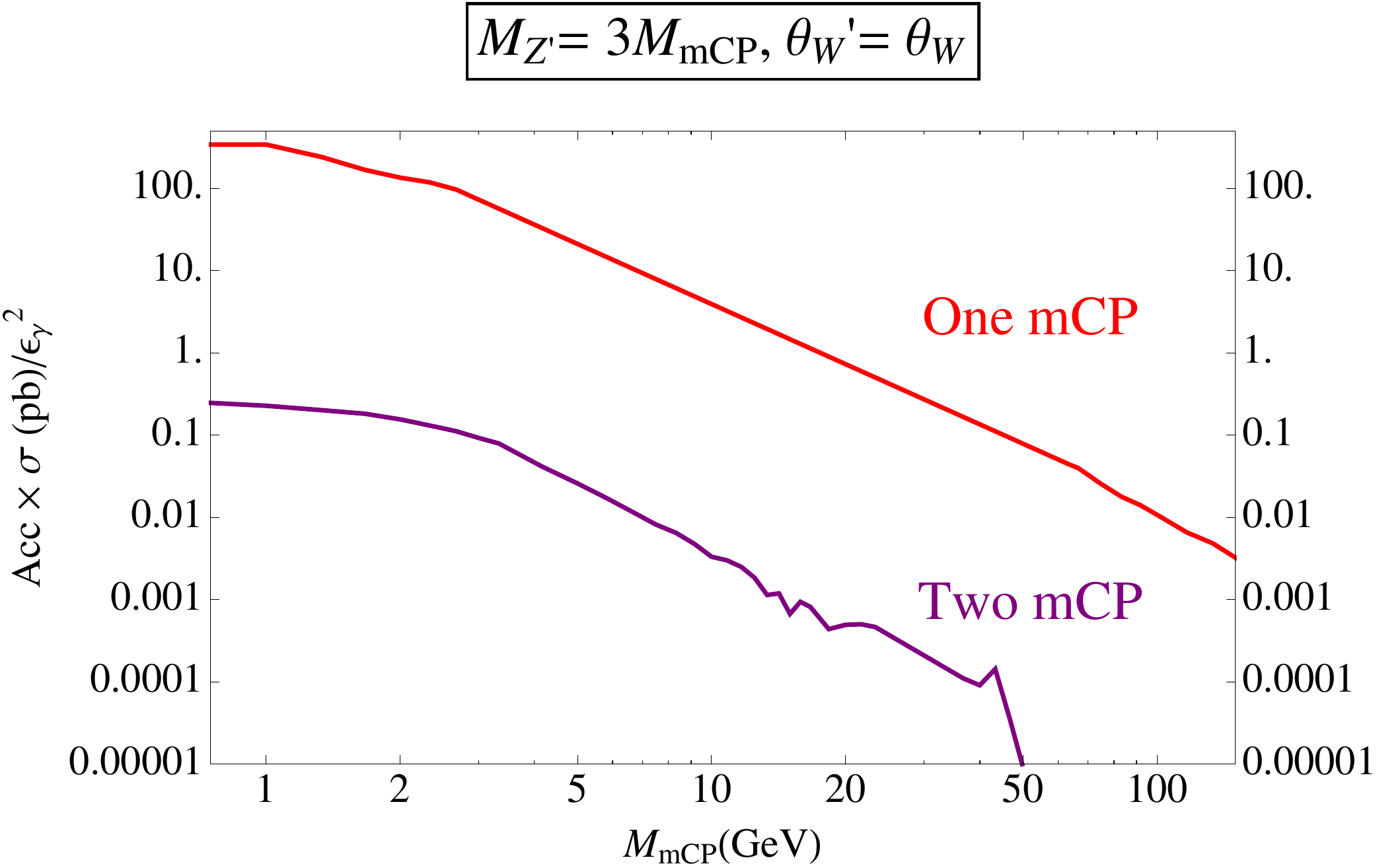}\\ 
\includegraphics[width=0.48 \textwidth ]{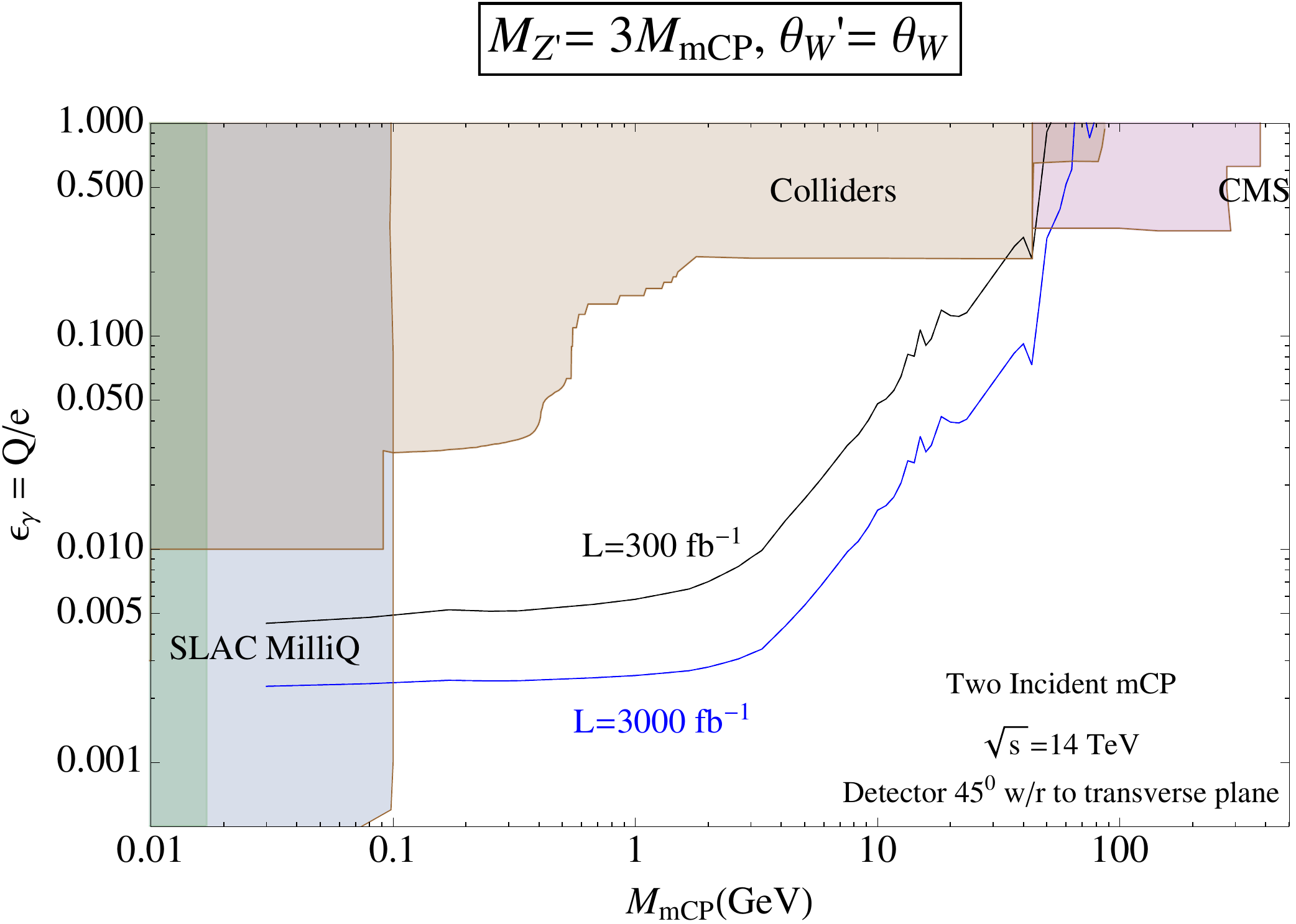}
\caption{We illustrate the new signature that opens up in the mixed phase. On top, we show the acceptance for one (as proposed in Ref.~\cite{Haas:2014dda} for the massless phase) mCP entering the detector deployed nearby the LHC main detectors. For the mixed phase, the massive $Z'$ can be produced in association with a jet with a smaller yet sufficient rate and acceptance, illustrated by the line for two incoming mCPs. The bottom panel shows the expected sensitivity when requiring two mCPs enter the detector instead of one.}
\label{fig:twovsone}
\end{figure}

The mixed phase of a DS that kinetically mixes with the SM offers additional possibilities for future striking signals. Here we comment on two possibilities. First, for $Z'$ produced in association with a jet, the $Z'$ will be boosted accordingly to compensate for the jet's transverse momentum. For sufficiently boosted $Z'$ with energy $E$ near $E \sim (2 M_{Z'}/\theta_{\rm{det}})$ --- where $\theta_{\rm{det}}$ is the angular opening of the detector proposed by Ref.~\cite{Haas:2014dda} --- {\it both} of the mCPs from the decay of the $Z'$ could enter the detector. This leads to the intriguing possibility of two simultaneous coincident hits in different bars of the detector. Additionally, such a signal could be correlated with a jet produced in association with the $Z'$ in the reaction $pp\rightarrow Z' + j$. The upper pane of Fig.~\ref{fig:twovsone} shows the cross section times acceptance for the case where one demands one mCP giving one or more PEs in each of the back-to-back scintillating bars versus the case where both mCPs enter the detector and deposit such a signal. While there is a considerable penalty in rate from requiring two mCPs entering the detector as opposed to one, the sensitivity does not degrade significantly as we show in the lower pane of Fig.~\ref{fig:twovsone}. The strong sensitivity to the two-incident-mCP scenario is still possible because what sets the sensitivity floor to the one-mCP scenario is not the signal rate, but the detection probability of one PE (see (\ref{eq:averagepe})), which decreases exponentially below $\epsG=2\times10^{-3}$. The two-mCP possibility could be instrumental should there arise an excess that is consistent with a charge as small as few-$10^{-3}e$. Interestingly, if such a signal is detected, the transverse momentum of the associated jet together with the opening angle of the mCP pair in the far-detector can be used to obtain a rough estimate of the mass of the $Z'$ through the relation above $E \sim (2 M_{Z'}/\theta_{\rm{det}})$.

The second possibility that is unique to the mixed phase arises in the case of large $e'/\sWp$ and sufficiently light $Z'$ and mCP. In this case, analogously to a QCD shower, the mCPs produced during the proton proton collisions could radiate $Z'$s which in turn decay back to mCPs, giving rise to a modest ``milli-jet''. Indeed, such a scenario could produce not one, but several mCPs entering through the detector. Finally, if the DS contains an additional gauge group that is confining at sufficiently low scales, then the mesons and baryons of that group may be the mCPs and a real ``milli-jet" may result.

\section{Discussion and Conclusions}
\label{sec:conclusions}

In this work we have largely ignored the issue of dark matter and have treated it mainly as a motivation to think of other sectors beyond the SM. But, it is only natural to wonder whether the model we discussed in this paper can yield a viable cosmological relic with sufficient abundance to explain the astronomical data. First, we note that particles in the DS that are charged under the DS photon cannot have a large cosmological abundance. This is so because these particles are also milli-charged under electromagnetism, see Eqs.~(\ref{eqn:couplings_of_photon}), and the relic abundance of mCPs with charge of $\gtrsim 10^{-3}$ as considered in this paper is strongly constrained to be $\Omega_{\rm mcp} h^2 < 0.001$ by the acoustic peaks in the cosmic microwave background~\cite{Dubovsky:2003yn,Dolgov:2013una}. To be sure, it is not difficult to satisfy this constraint since the pair annihilation of mCPs into DS photons has a large cross-section\footnote{The non-relativistic cross-section is $\sigma v = \pi (\alpha')^2/\MQ^2$, which would keep the mCPs in thermal equilibrium with the DS photon bath and deplete the mCP relic abundance to the level of $\Omega_{\rm mcp} h^2 \approx 10^{-5}\left(\MQ/\GeV \right)^2 \left(10^{-2}/\alpha' \right)^2$.}, but it does mean that such mCPs cannot be the dark matter. 

However, not all the particles in the DS have to be charged under the DS photon. For example, DS neutrino-like particles that are only charged under the $Z'$ would not acquire a milli-charge under electromagnetism. Instead their contact with the SM is only through the $Z'$ and the $Z$ vector bosons, see Eqs.~(\ref{eqn:Zcouplings}) and (\ref{eqn:Zpcouplings}), and is thus suppressed by $\kappa \sWp \sim 10^{-3}$. This makes it easier to accommodate a larger relic abundance of DS neutrino-like particles (which we collectively refer to as WIMPs in what follows), but this possibility is not without its constraints, as we now briefly discuss. 

One possibility is the secluded regime of Ref.~\cite{Pospelov:2007mp} where the WIMP mass is several hundred GeVs, and $\mZp \sim 100\GeV$. The correct relic abundance can then be obtained with $g_2^{\prime 2}/4\pi \sim 10^{-2}$ through the WIMP's pair annihilation into a pair of $Z'$. With $\kappa \sWp \sim 10^{-3}$ this easily evades all the collider bounds, but it can perhaps be discovered in future direct detection experiments with sensitivity to WIMP-nucleon cross-section of $\sigma_n \sim 10^{-45}-10^{-46}\cm^2$. 

Another possibility is of light WIMP with mass $\lesssim \GeV$, but where the direct annihilation into a pair of $Z'$ is kinematically forbidden. The strongest constraints on this scenario come from early universe energy injection during hydrogen recombination that distorts the cosmic microwave background~\cite{Padmanabhan:2005es,Slatyer:2009yq}. These constraints, which grow at lower DM mass and are especially strong below 10 GeV, exclude the possibility of a thermal freeze-out through off-shell $Z'$ annihilation into SM charged particles. However, if there is a strong particle - anti-particle asymmetry in the DS then this constraint may be avoided as discussed in Ref.~\cite{Lin:2011gj}. 

The above examples illustrate that the cosmological dark matter may be connected to neutral particles of the model we considered in this work. But, given how uncertain the structure of the DS is, if it exists at all, we prefer not to speculate further on the issue beyond these representative examples. Ultimately the strength of the search described in this paper and Ref.~\cite{Haas:2014dda} is that it is independent of any assumptions regarding the relic abundance associated with the DS and instead directly probes the existence of mCPs. 

In the present paper we explored a general theory with mCPs that continuously interpolates between the Holdom phase and the Okun phase of theories with a vector boson that kinetically mixes with the SM hypercharge. The main phenomenological signatures of this model are the appearance of mCPs as well as a massive vector boson, the $Z'$, that couples to the electromagnetic current but predominantly decays into mCPs. The model is only weakly constrained when the mass scale associated with these new particles is above about 100 MeV. We explored the sensitivity of a recently proposed dedicated experiment at the LHC~\cite{Haas:2014dda} to this model in general (Fig.~\ref{fig:moneyplotlowthetap}), and to the resonant production of mCPs through the $Z'$ in particular (Fig.~\ref{fig:twovsone}).

\section*{Acknowledgments}
We thank the Abdus Salam International Center for Theoretical Physics and Fermilab for their hospitality. We would like to thank Andy Haas and Chris Hill for useful discussions. We also thank Roni Harnik for very stimulating discussions in the early stages of this project.  Research at Perimeter Institute is supported by the Government of Canada through Industry Canada and by the Province of Ontario through the Ministry of Research and Innovation. EI is partially supported by the Ministry of Research and Innovation - ERA (Early Research Awards) program. IY is supported in part by funds from the NSERC of Canada. 

\bibliography{milli_world}
\end{document}